\documentclass[twocolumn,preprintnumbers,prb,amssymb,superscriptaddress,floatfix]{revtex4}
\usepackage{xcolor,graphicx}
\usepackage{bm}
\usepackage{amssymb}
\usepackage{amsmath} 
\usepackage{units}

\begin{document}

\title{Octahedral rotation patterns in strained EuFeO$_3$ and other $\pmb{Pbnm}$ perovskite films: Implications for hybrid improper ferroelectricity}
\author{A.\ K.\ Choquette}
\author{C.\ R.\ Smith}  
\affiliation{
Department of Materials Science and Engineering, Drexel University, Philadelphia, PA 19104\\}
\author{R.\ J.\ Sichel-Tissot}
\affiliation{
Department of Materials Science and Engineering, Drexel University, Philadelphia, PA 19104\\}
\affiliation{
Materials Science Division, Argonne National Laboratory, Argonne, IL 60439\\}
\author{E.\ J.\ Moon}
\author{M.\ D.\ Scafetta}
\affiliation{
Department of Materials Science and Engineering, Drexel University, Philadelphia, PA 19104\\}
\author{E. Di Gennaro}
\author{F. Miletto Granozio}
\affiliation{
CNR-SPIN and Dipartimento di Fisica, Universit\`{a} di Napoli ``Federico II'', 80126 Naples, Italy}
\author{E.\ Karapetrova}
\affiliation{
X-ray Science Division, Argonne National Laboratory, Argonne, IL 60439\\}
\author{S.\ J.\ May}
\email{smay@coe.drexel.edu}
\affiliation{
Department of Materials Science and Engineering, Drexel University, Philadelphia, PA 19104\\}

\date{\today}


\begin{abstract}
We report the relationship between epitaxial strain and the crystallographic orientation of the in-phase rotation axis and $A$-site displacements in $Pbnm$-type perovskite films. Synchrotron diffraction measurements of EuFeO$_3$ films under strain states ranging from 2~\% compressive to 0.9~\% tensile on cubic or rhombohedral substrates exhibit a combination of $a^-a^+c^-$ and $a^+a^-c^-$ rotational patterns.  We compare the EuFeO$_3$ behavior with previously reported experimental and theoretical work on strained $Pbnm$-type films on non-orthorhombic substrates, as well as additional measurements from LaGaO$_3$, LaFeO$_3$, and Eu$_{0.7}$Sr$_{0.3}$MnO$_3$ films on SrTiO$_3$. Compiling the results from various material systems reveals a general strain dependence in which compressive strain strongly favors $a^-a^+c^-$ and $a^+a^-c^-$ rotation patterns and tensile strain weakly favors $a^-a^-c^+$ structures. In contrast, EuFeO$_3$ films grown on $Pbnm$-type GdScO$_3$ under 2.3~\% tensile strain take on a uniform $a^-a^+c^-$ rotation pattern imprinted from the substrate, despite strain energy considerations that favor the $a^-a^-c^+$ pattern. These results point to the use of substrate imprinting as a more robust route than strain for tuning the crystallographic orientations of the octahedral rotations and $A$-site displacements needed to realize rotation-induced hybrid improper ferroelectricity in oxide heterostructures.
\end{abstract}

\maketitle

\section{Introduction}

Epitaxial heterostructures of $AB$O$_3$ perovskite oxides have attracted considered interest as a route toward altering or enhancing properties through epitaxial strain, superlattice formation, and interfacial phenomena.\cite{Schlom2008, Hwang2012, Zubko2011,Bhattacharya14} Recently, the control of local atomic structure, in particular $B$O$_6$ octahedral distortions and rotations and $A$-site displacements, has emerged as a promising strategy for designing functional properties in perovskite films.\cite{Bousquet08,Rondinelli2012a,BenedekJSSC,MoonNC} One example of structure-driven design in oxide heterostructures is the prediction of hybrid improper ferroelectricity in ($A^{'}B$O$_3$)/($AB$O$_3$) superlattices where both $A^{'}B$O$_3$ and $AB$O$_3$ are perovskites that exhibit the orthorhombic $Pbnm$ structure in bulk.\cite{Rondinelli2012, Mulder2013, GhoshPRB15, Benedek15} In such superlattices, the inequivalent displacements of the $A$ and $A^{'}$ cations produce a ferrielectric state. This design principle is predicated on the $A$-site displacements occuring within the plane of the superlattice, perpendicular to the superlattice growth direction. A similar design approach has been used to predict that (SrRuO$_3$)$_1$/(CaRuO$_3$)$_1$ superlattices are polar metals.\cite{Puggioni14}

A key challenge for experimentally verifying such predictions lies in the quantitative measurement of octahedral behavior and $A$-site positions in thin films, as the primary technique used in bulk perovskites - powder diffraction - is not accessible in studies of epitaxial films. Recent work has shown the promise of synchrotron diffraction,\cite{May2010,ChangPRB11,Rotella2012,Johnson13,LuPRB13,Fister14,ZhangPRB14,Zhai14,Biegalski14} coherent Bragg rod analysis,\cite{Fister14,Kumah14} electron microscopy,\cite{JiaPRB09,BorisevichPRL10,AsoCGD14} and electron diffraction\cite{HwangAPL12} to probe octahedral rotations in perovskite films. In particular, the synchrotron diffraction approach is based on the measurement of half-order Bragg peaks that arise from the unit cell doubling nature of the octahedral rotations.\cite{Glazer1975,Woodward05} The presence and absense of specific half-order peaks is a direct signature of the pattern of octahedral rotations within the material. The rotation pattern is denoted using Glazer notation, in which in-phase, out-of-phase, or absense of rotations are signified by +, -, or 0 superscripts, respectively, along a given pseudocubic direction.\cite{Glazer1972,Woodward1997} Axes with equal rotational magnitude are denoted by the same letter. For example, the $a^-a^-a^-$ pattern consists of equal out-of-phase rotation angles along all three pseudocubic axes, while the $a^-a^-c^+$ pattern has equivalent out-of-phase rotations along two axes and an in-phase rotation of differing magnitude along the third axis. This latter pattern corresponds to the orthorhombic $Pbnm$ perovskite structural variation, which is one of the most common crystal structures for oxide perovskites.\cite{Thomas1989,Woodward1997a} Materials in this structure also exhibit $A$-site displacements in the plane normal to the in-phase rotation axis.

There is limited understanding of what determines the direction of the in-phase rotation axis, and therefore the $A$-site displacements, in epitaxial $Pbnm$-type perovskite films and superlattices despite the clear importance of this knowledge for the realization of new ferroic materials. While there have been numerous reports of the rotation pattern within a single film,\cite{Copie2013,Proffit2008,Choi2010a,Han2009,Kan2013} systematic experimental studies probing the effect of a single variable, such as strain or composition, on the rotation pattern in $Pbnm$-type films are lacking.

In this work, we report on the rotation patterns of strained EuFeO$_3$ films, a perovskite that exhibits the $Pbnm$ structure in bulk form.\cite{Marezio1970a} These results are compared with previously reported experimental and theoretical work on strained $Pbnm$-type films, as well as new measurements from LaGaO$_3$, LaFeO$_3$, and Eu$_{0.7}$Sr$_{0.3}$MnO$_3$ films, revealing a general strain dependence in which compressive strain strongly favors $a^-a^+c^-$ and $a^+a^-c^-$ rotation patterns and tensile strain weakly favors $a^-a^-c^+$ structures. However, EuFeO$_3$ grown on orthorhombic GdScO$_3$ (110) exhibits a uniform $a^-a^+c^-$ orientation matching that of the substrate, despite the 2.3~\% tensile strain imposed by the substrate. This result indicates that the use of substrate templating is a more deterministic route than strain for controlling the in-phase rotation axis and $A$-site displacement orientation in perovskite films and superlattices.

\section{Experimental Techniques}

EuFeO$_3$ films were grown on SrTiO$_3$ (STO) (001), (LaAlO$_3$)$_{0.3}$(Sr$_2$AlTaO$_6$)$_{0.7}$  (LSAT) (001), LaAlO$_3$ (LAO) (001), and GdScO$_3$ (GSO) (110) substrates using oxide molecular beam epitaxy (MBE). The growth conditions are described in Ref.~\onlinecite{Choquette2015}. The thickness of the EuFeO$_3$ films are between 35-40 unit cells (13-15~nm) thick as determined from x-ray diffraction. LaFeO$_3$ and Eu$_{0.7}$Sr$_{0.3}$MnO$_3$ films were also deposited by MBE on STO (001) using conditions reported in Ref. \onlinecite{Scafetta13} and Ref. \onlinecite{Moon2014a}, respectively. The LaGaO$_3$ film was grown on STO (001) using pulsed laser deposition as described in Ref. \onlinecite{Perna10}. The LaGaO$_3$, Eu$_{0.7}$Sr$_{0.3}$MnO$_3$, and LaFeO$_3$ films are 25, 40, and 129 unit cells thick, respectively. A previously published reciprocal space map from this LaFeO$_3$ film confirms that it is coherently strained to the STO substrate.\cite{Scafetta13} Synchrotron diffraction measurements were performed at Sector 33-BM-C of the Advanced Photon Source. All measurements were carried out at room temperature. Photon energies of 15.5~keV and 16~keV were used for the EuFeO$_3$ and Eu$_{0.7}$Sr$_{0.3}$MnO$_3$ measurements, respectively. The LaGaO$_3$ and LaFeO$_3$ films were measured with 10~keV photons. The GenX software package\cite{Bjorck2007} was used to simulate the measured ($00L$) data, from which $c$-axis parameters and film thicknesses were obtained.  Volume fractions of different structural orientations were obtained by analyzing peak areas after applying Lorentz polarization and beam footprint corrections.

\section{E\lowercase{u}F\lowercase{e}O$_3$ Films}
  
\begin{figure}
\includegraphics[width=3.5 in]{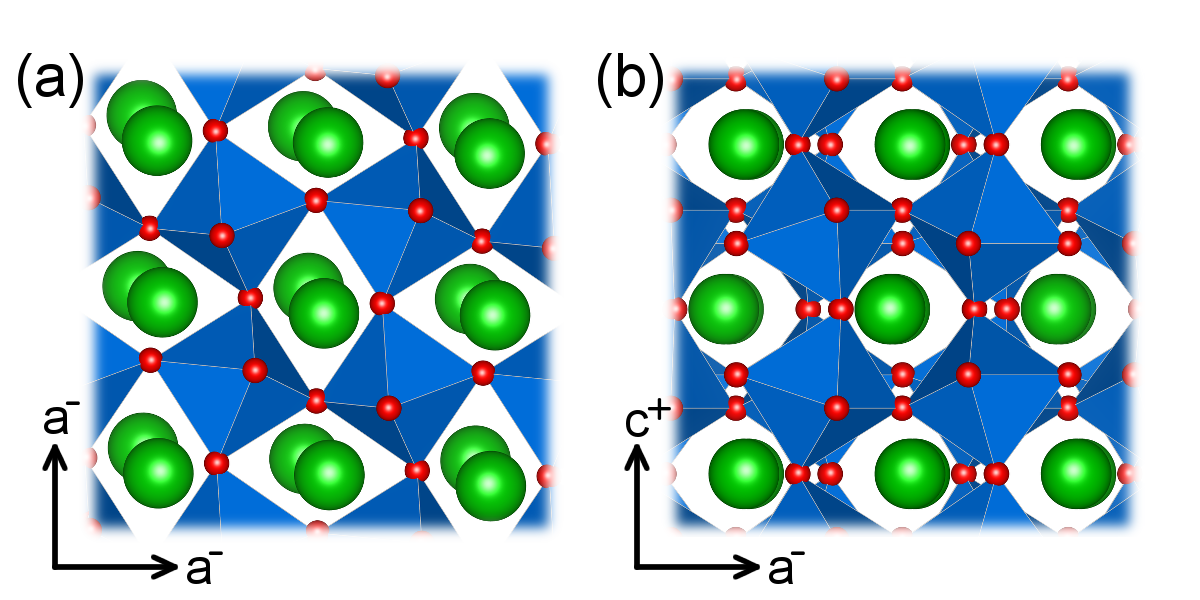}
\caption{(Color online) The crystal structure of bulk EuFeO$_3$, representative of $Pbnm$-type perovskites, viewed along the pseudocubic [001] (a) and [100] (b) directions; structural data from Ref.~\onlinecite{Marezio1970a}.}
\label{fig:Fig1}
\end{figure}
  
  \begin{figure*}
\includegraphics[width=7 in]{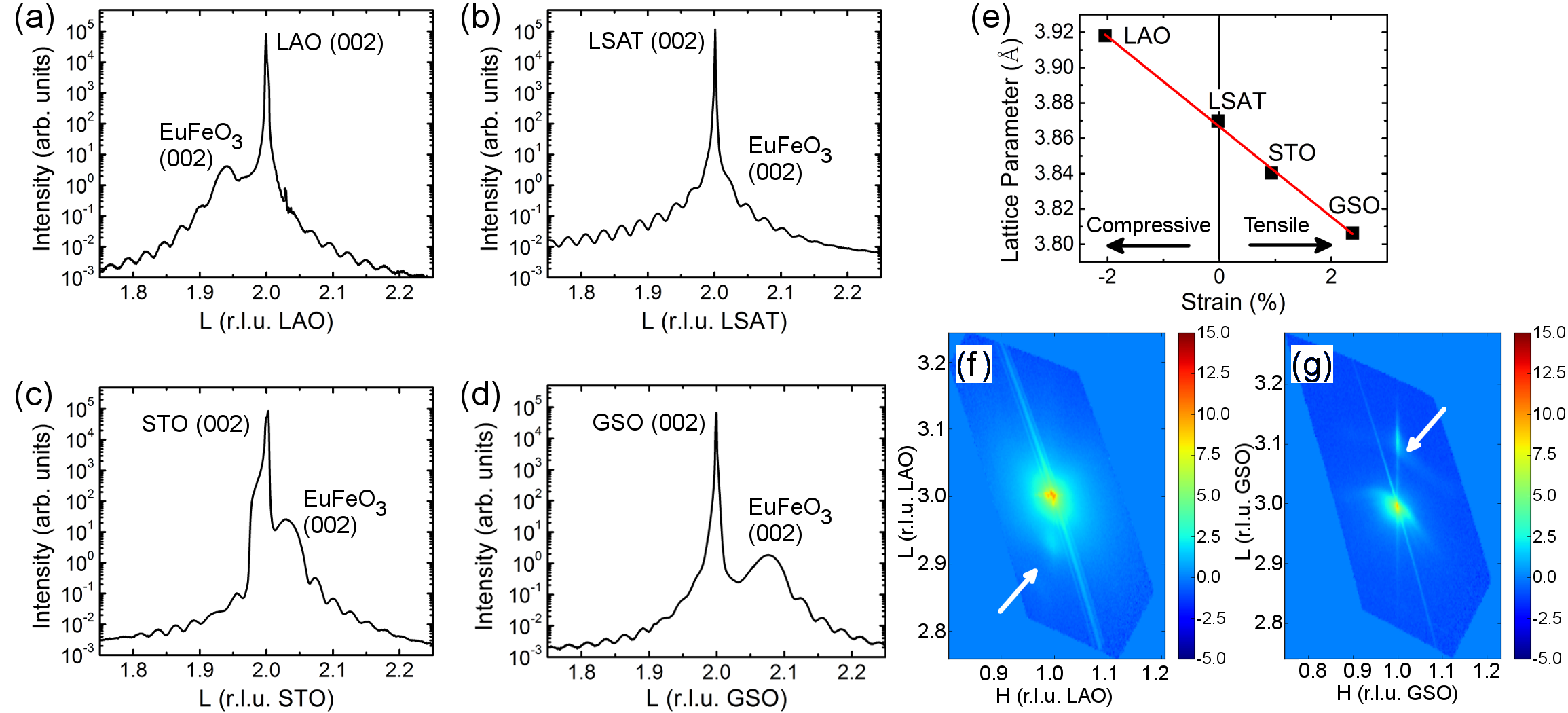}
\caption{(color online) Scans along $00L$ through the (002) peak for EuFeO$_3$ films on LAO (a), LSAT (b), STO (c) and GSO (d). The $c$-axis parameter as a function of in-plane strain is shown in (e). The red line is a guide for the eye. Reciprocal space maps about the (113) peak for the EuFeO$_3$ film on LAO (f) and GSO (g). Arrows highlight the Bragg peak from the EuFeO$_3$ films. The scale bars in (f) and (g) indicate the natural log of the measured intensity.}
\label{fig:Fig2}
\end{figure*}

  \begin{figure}
\includegraphics[width=3.0 in]{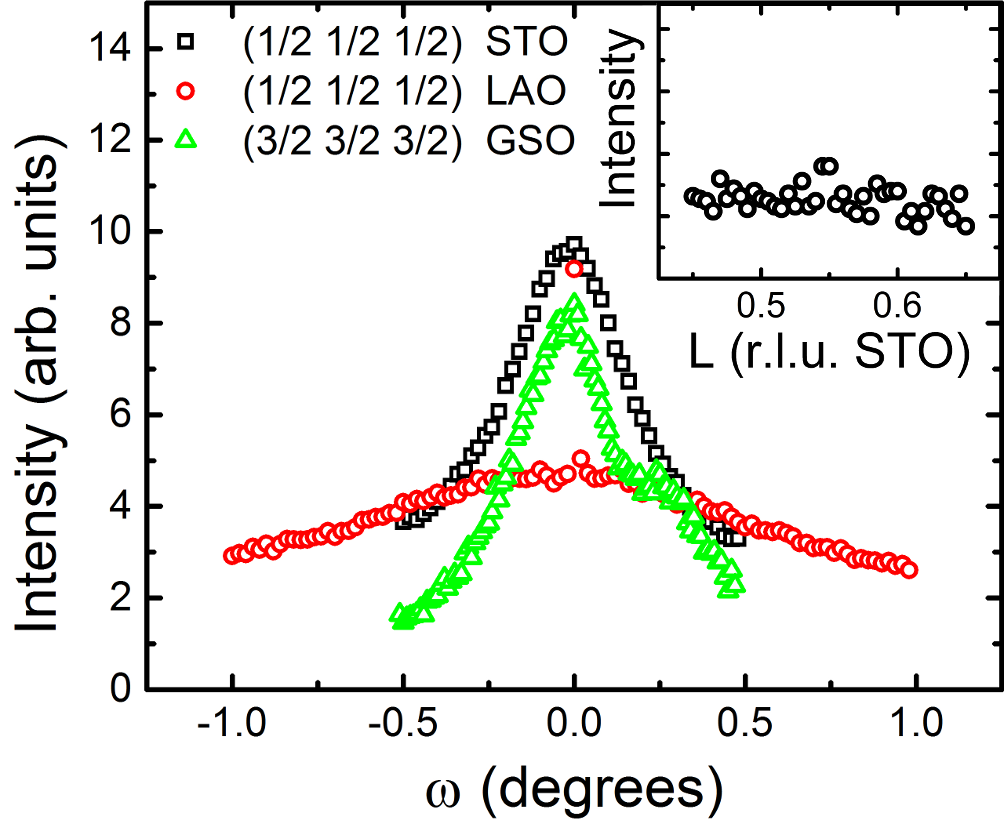}
\caption{(color online)  Omega scans through $H$=$K$=$L$ half-order diffraction peaks from EuFeO$_3$ films on STO, LAO and GSO. The inset shows an $L$-scan through the ($\frac{1} {2}$ $\frac{1} {2}$ $\frac{1} {2}$) condition for a LaNiO$_3$ film on STO.}
\label{fig:Fig3}
\end{figure}

  \begin{figure*}
\includegraphics{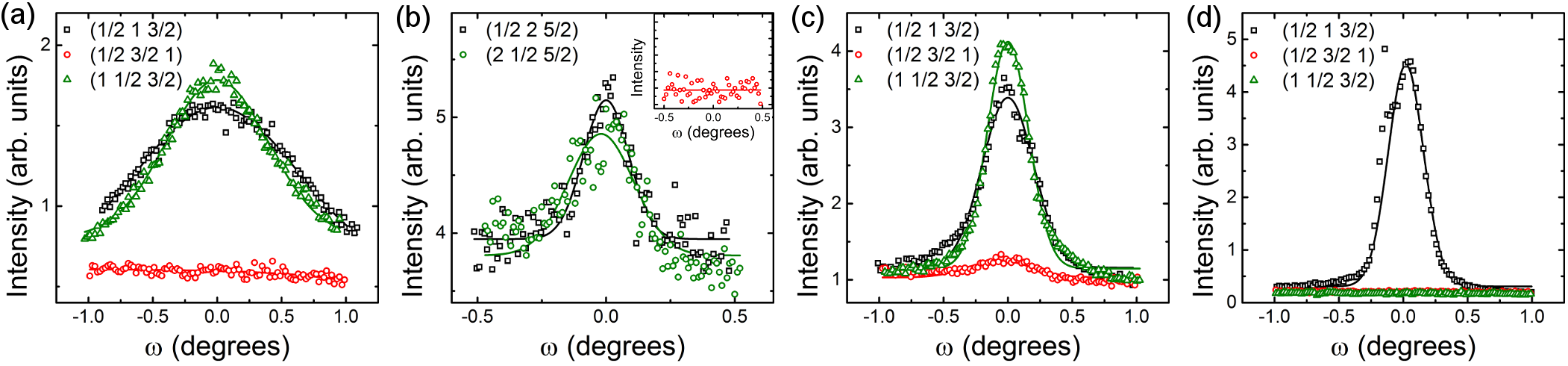}
\caption{(color online) Omega scans through half-order peaks where the integer index indicates the orientation of the in-phase rotation axis. The EuFeO$_3$ films on (a) LAO  and (b) LSAT both show evidence of mixed $a^+a^-c^-$ and $a^-a^+c^-$ domain structure. The inset of (b) is an omega scan through the ($\frac{1} {2}$ $\frac{5} {2}$ 2) condition for the EuFeO$_3$ film on LSAT. The film on (c) STO exhibits a mixed $a^+a^-c^-$ and $a^-a^+c^-$ domain population with a small fraction of $a^-a^-c^+$ domains.  The film on (d) GSO shows only the $a^-a^+c^-$ structural orientation.}
\label{fig:Fig4}
\end{figure*}

  \begin{figure}
\includegraphics{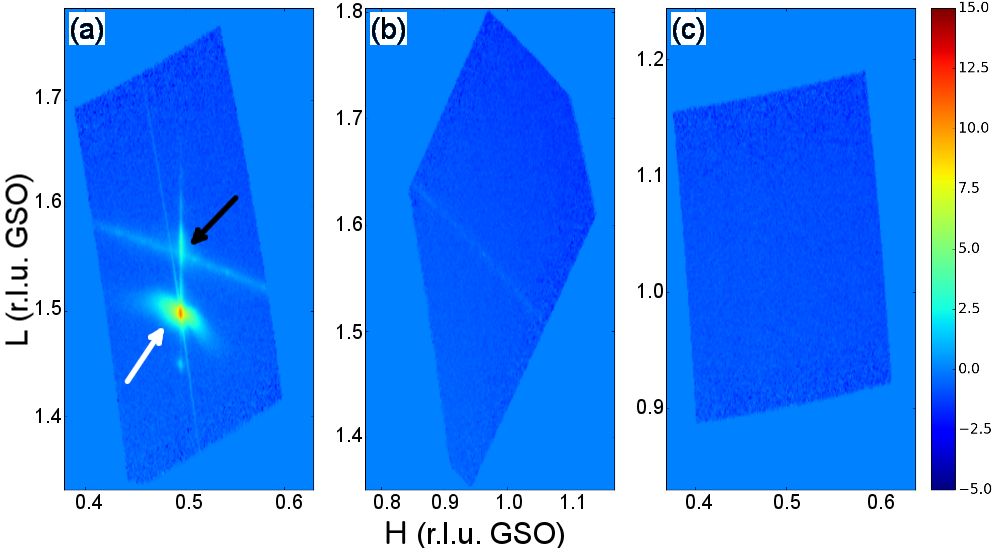}
\caption{(color online) Regions of reciprocal space near the \{$\frac{1} {2}$ $\frac{3} {2}$ 1\} Bragg conditions for EuFeO$_3$ on GSO. The ($\frac{1} {2}$ 1 $\frac{3} {2}$) peak is shown in (a); the (1 $\frac{1} {2}$ $\frac{3} {2}$) region of reciprocal space is shown in (b); the ($\frac{1} {2}$ $\frac{3} {2}$ 1) region of reciprocal space is shown in (c).  Only the ($\frac{1} {2}$ 1 $\frac{3} {2}$) peak is present (a), mirroring the substrate. In (a), the white arrow highlights the peak from GSO; the black arrow highlights the peak from the EuFeO$_3$. The scale bar indicates the natural log of the measured intensity.}
\label{fig:Fig5}
\end{figure}

  \begin{figure}
\includegraphics[width=2.5 in]{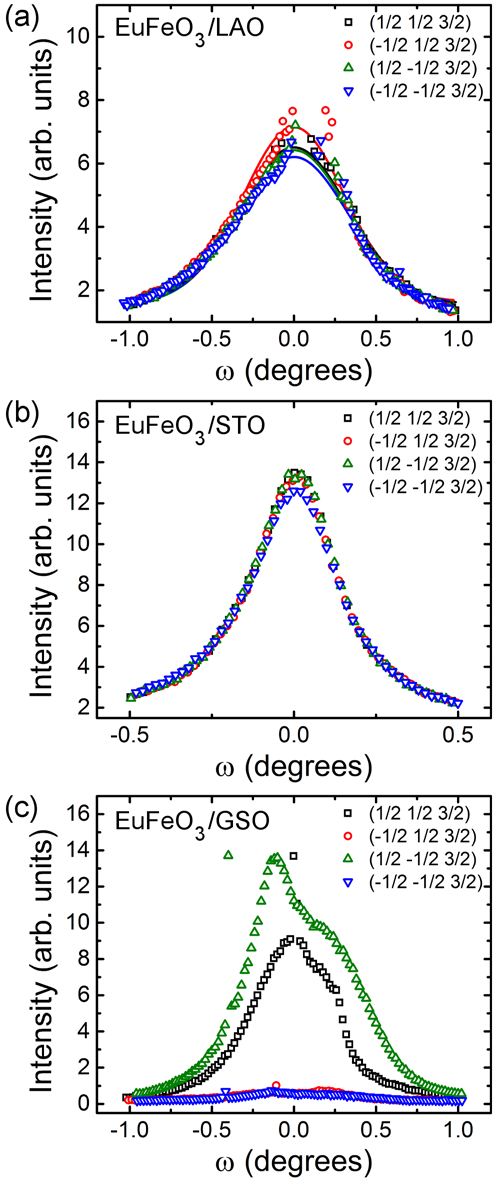}
\caption{(color online)  The ($\pm\frac{1} {2}$ $\pm\frac{1} {2}$ $\frac{3} {2}$) peaks measured from EuFeO$_3$ films on (a) LAO, (b) STO, and (c) GSO.  The films on LAO and STO show evidence of equal populations of the rotational domains. The film on GSO shows evidence of unequal rotational domains.}
\label{fig:Fig6}
\end{figure}

The orthorhombic $Pbnm$ structure is one of the most common perovskite variants among oxides and is the structure exhibited by bulk EuFeO$_3$ at room temperature. Within the thin film community, the orthorhombic lattice is commonly converted to a pseudocubic structure in which the orthorhombic [100] is equivalent to the pseudocubic [110] and $a_o = \sqrt{2} a_p$. The [001] direction is unchanged but the pseudocubic $c$-axis parameter is half that of the orthorhombic $c$-axis parameter. This pseudocubic lattice will be used throughout this work. Two key features of this structure, shown in Fig.~\ref{fig:Fig1}, are the presence of the $a^-a^-c^+$ rotation pattern and $A$-displacements. The $a^-a^-c^+$ rotation pattern indicates out-of-phase $B$O$_6$ rotations about two in-plane directions (pseudocubic [100] and [010]) and in-phase rotations along the out-of-plane [001] direction. The $A$-site cations are displaced within the plane perpendicular to the $c^+$ rotation axis along directions close to the \textless110\textgreater. Both the octahedral rotations and $A$-site displacements act to double the pseudocubic unit cell, leading to half-order diffraction peaks. Throughout this work, we will denote the growth direction as the $c$-axis of the film. The $a^-a^+c^-$ pattern indicates the in-phase axis lies along an in-plane film direction. The $a^-a^-c^+$ pattern indicates that the in-phase axis lies along the out-of-the-plane film direction (the growth direction). In this study, we do not quantify the rotation angles and therefore cannot distinguish between $a^-a^+c^-$ and $a^-b^+c^-$. We use $a^-a^+c^-$ throughout this work with the understanding that this entails both $a^-a^+c^-$ and $a^-b^+c^-$. However, for films grown on STO, LAO, and LSAT substrates, we anticipate that the in-plane rotations, $\alpha$ and $\beta$, are equal due to the same in-plane lattice constants along the $a$ and $b$-axes. For films on orthorhombic GSO, inequivalent $\alpha$ and $\beta$ angles would be expected leading to $a^-b^+c^-$ or $a^-b^-c^+$ patterns because the in-plane lattice constants of the substrate are not identical.

The bulk structure of EuFeO$_3$ has previously been determined from powder diffraction measurements.\cite{Marezio1970a}  It has pseudocubic lattice parameters, taken from the Fe-Fe distances of 3.882 \AA~ along the $a^-$ axes and 3.842 \AA  ~along the $c^+$ axis (the orthorhombic long axis). The reduced $B$-$B$ distance along the in-phase axis compared to the out-of-phase axes is a common feature of the $Pbnm$ structure, having also been reported in bulk LaFeO$_3$,\cite{Falcon97} LaGaO$_3$,\cite{Vasylechko99}, LaTiO$_3$,\cite{Komarek07} LaVO$_3$,\cite{Bordet93} LaCrO$_3$,\cite{Tezuka98} CaTiO$_3$,\cite{Liu93} CaFeO$_3$,\cite{Takeda00} and CaMnO$_3$.\cite{Chmaissem01} Based solely on these distances, one would expect the in-phase axis to lie out-of-the-plane ($a^-a^-c^+$) for films under tensile strain and in-the-plane ($a^+a^-c^-$ or $a^-a^+c^-$) for films under compressive strain in order to minimize the lattice mismatch with the substrate. 

EuFeO$_3$ films were deposited on a variety of commercially available substrates. The lattice mismatch between EuFeO$_3$ and the substrates leads to an average 2\% compressive strain on LAO (lattice parameter 3.791 \AA), \textless 0.1\% strain on LSAT (3.868 \AA), 0.9\% tensile strain on STO (3.905 \AA), and 2.3\% tensile strain on GSO (3.968 \AA).  The measured $00L$ scans are shown in Fig.~\ref{fig:Fig2}(a-d). The obtained EuFeO$_3$ $c$-axis parameters are 3.918 \AA~ on LAO, 3.840 \AA~ on STO, 3.869 \AA~ on LSAT, and 3.806 \AA~ on GSO, consistent with the strain states of the films, shown in Fig.~\ref{fig:Fig2}(e).  Further verification of the strain state comes from reciprocal space maps measured about the (113) peak, as shown Fig. 2(f) for EuFeO$_3$/LAO and Fig. 2(g) for EuFeO$_3$/GSO. The Bragg peak from the films occurs at the same $H$ and $K$ values as that of the substrate, indicating that the films are coherently strained.  Similar reciprocal space maps for the films on LSAT and STO were previously reported in Ref.~\onlinecite{Choquette2015}, which confirm the EuFeO$_3$ films are strained to LSAT and STO. 

A broad survey of half-order diffraction peaks were measured to determine the pattern of octahedral rotations in the films. We first present measurements at $H=K=L$ conditions. These peaks arise from $A$-site displacements with minimal intensity contribution from octahedral rotations.\cite{Glazer1975} Figure~\ref{fig:Fig3} displays $\omega$ scans, as commonly referred to as rocking curves, through $H=K=L$ regions of reciprocal space for EuFeO$_3$ films on STO, LAO and GSO, which all exhibit peaks. The presence of these peaks is consistent with the presence of $A$-site displacements in the $Pbnm$-type structure. In contrast, a 45 u.c. thick LaNiO$_3$ film, shown inset of Fig. 3, does not exhibit a ($\frac{1} {2}$ $\frac{1} {2}$ $\frac{1} {2}$) peak as expected for a $R\bar3c$-type perovskite lacking $A$-site displacements. The broad and intense ($\frac{1} {2}$ $\frac{1} {2}$ $\frac{1} {2}$) peak from the LSAT substrate\cite{Sang2015} prevented measurement of the film peak at this condition for the EuFeO$_3$/LSAT sample.

We next move to Bragg conditions in which one of the reciprocal lattice positions is an integer and the other two are unequal half-order positions, for example ($\frac{1} {2}$ 1 $\frac{3} {2}$) or ($\frac{1} {2}$ 2 $\frac{5} {2}$) where $K$ is an integer and $H$ $\neq$ $L$. Figure 4 shows a series of three peaks in which either \textit{H}, \textit{K}, or \textit{L} is an integer, and the total momentum transfer, \textit{q}, is kept approximately constant. These peaks are present only when the integer reciprocal lattice variable is parallel to the real space direction of the in-phase rotation axis.\cite{Glazer1975} For example, an $a^-a^-c^+$ pattern produces a ($\frac{1} {2}$ $\frac{3} {2}$ 1) peak.  The \textit{A}-site displacements perpendicular to the direction of the in-phase rotation also contribute intensity to these peaks. Therefore, the presence of a ($\frac{1} {2}$ $\frac{3} {2}$ 1)-type peak allows for the orientation of the in-phase rotation axis to be determined. For the  films on LAO [Fig.~\ref{fig:Fig4}(a)] and on LSAT [Fig.~\ref{fig:Fig4}(b)], peaks with an integer in either $H$ or $K$ are observed, while peaks with an integer $L$ value are absent. Figure 4(a) shows a series of (1 $\frac{1} {2}$ $\frac{3} {2}$)-type peaks for the EuFeO$_3$/LAO sample. While the ($\frac{1} {2}$ 1 $\frac{3} {2}$) and (1 $\frac{1} {2}$ $\frac{3} {2}$) have approximately equal intensity, no intensity is measured at the ($\frac{1} {2}$ $\frac{3} {2}$ 1). Similar data is obtained from EuFeO$_3$/LSAT, shown in Fig. 4(b), where a larger value of $L$ is used to better separate the film and substrate peaks. For the EuFeO$_3$/STO film, shown in Fig.~\ref{fig:Fig4}(c), the majority of the film takes a structure of $a^+a^-c^-$ or $a^-a^+c^-$, with only a small fraction (4\%) of the film exhibiting $a^-a^-c^+$, as has previously been reported.\cite{Choquette2015} 

In contrast, this multi-domain trend is not observed in the EuFeO$_3$/GSO film. Instead, the film exhibits a uniform $a^-a^+c^-$ pattern, which matches that of the GSO substrate. As shown in Fig.~\ref{fig:Fig4}(d), only the ($\frac{1} {2}$ 1 $\frac{3} {2}$) peak is present and both the (1 $\frac{1} {2}$ $\frac{3} {2}$) and ($\frac{1} {2}$ $\frac{3} {2}$ 1) peaks are absent. The ($\frac{1} {2}$ 1 $\frac{3} {2}$) peak is asymmetric due to some contribution from the substrate in the $\omega$ scan. $L$ scans through these three regions of reciprocal space are presented as supplemental materials (Fig. S1).\cite{note}

Reciprocal space maps measured near these same set of peaks further demonstrate that the film rotation behavior is dependent on that of the substrate. For the film on GSO, peaks at ($\frac{1} {2}$ 1 $\frac{3} {2}$) from both the substrate and film can be seen in Fig.~\ref{fig:Fig5}(a), with the white and black arrows highlighting the substrate and film peak, respectively. There is no intensity from either the substrate or film at the (1 $\frac{1} {2}$ $\frac{3} {2}$) and ($\frac{1} {2}$ $\frac{3} {2}$ 1) conditions [Fig.~\ref{fig:Fig5}(b,c)], consistent with a uniform $a^-a^+c^-$ pattern in both the substrate and film.

The \{$\pm\frac{1} {2}$ $\pm\frac{1} {2}$ $\frac{3} {2}$\} series of peaks, shown in Fig. 6, provides additional evidence for the presence of mixed $a^+a^-c^-$ and $a^-a^+c^-$ patterns on LAO and STO, and uniform $a^-a^+c^-$ orientation on GSO. These peaks arise from out-of-phase rotations within the plane of the film ($a^-$)\cite{Glazer1975} and from $A$-site displacements within the plane perpendicular to the rotation axis. Therefore, the presence of these peaks indicates that the EuFeO$_3$/LAO and EuFeO$_3$/STO films are not $a^+a^+c-$ but instead contain regions of both $a^+a^-c^-$ and $a^-a^+c^-$ patterns. This is consistent with previous scanning transmission electron microscopy results obtained from an EuFeO$_3$/STO film in which $Pbnm$-type rotations were observed with the in-phase axis lying along different pseudocubic directions.\cite{Choquette2015} 

Additionally, within a given rotation pattern, different rotational domains can arise.  Each domain is defined by how the closest octahedron to the origin rotates (clockwise or counterclockwise) about each axis, which in turn dictates the displacement direction of the oxygen atoms within that rotation pattern. To probe these rotational domains, symmetrically equivalent half-order peaks with a fixed $L$ are measured.\cite{May2010} For the film on STO, we find that the intensity of the four \{$\pm\frac{1} {2}$ $\pm\frac{1} {2}$ $\frac{3} {2}$\} peaks are equal, indicating an equal population of the rotational domains as would be expected for growth on a cubic substrate. Similar data is obtained from the EuFeO$_3$/LAO sample, indicating that the rotational domains from LAO, which has an $a^-a^-a^-$ pattern, are not transferred into the film due to the symmetry mismatch at the interface. $L$ scans through these \{$\pm\frac{1} {2}$ $\pm\frac{1} {2}$ $\frac{3} {2}$\} peaks are shown in the supplemental materials (Fig. S2). This data clearly demonstrates that the rotational domain populations are not equal in the LAO substrate in contrast to the EuFeO$_3$ film, providing further evidence that the LAO is not imprinting rotational information into the film beyond the effect of strain. In contrast, the ($\frac{1} {2}$ $\frac{1} {2}$ $\frac{3} {2}$) and ($\frac{1} {2}$ -$\frac{1} {2}$ $\frac{3} {2}$) peaks are significantly more intense than the (-$\frac{1} {2}$ $\frac{1} {2}$ $\frac{3} {2}$) and (-$\frac{1} {2}$ -$\frac{1} {2}$ $\frac{3} {2}$) peaks in the EuFeO$_3$ film on GSO. As shown in supplemental Fig. S3, the same trend in peak intensities is found in the GSO substrate. This result indicates that not only is the rotation pattern imprinted from the GSO substrate, but the rotational domains within that pattern are also transferred from the substrate to film. 

\section{Other $\pmb{Pbnm}$-type Perovskites}

\subsection{Films on non-$\pmb{Pbnm}$ substrates}
Based on the purely geometric considerations described earlier in the text, one would expect that the in-phase rotation axis in $Pbnm$-type perovskite films on cubic substrates would depend on the epitaxial strain state.  A mixed $a^-a^+c^-$ and $a^+a^-c^-$ rotational pattern would be expected for compressive strain, putting the shorter pseudocubic in-phase axis in the plane of the film thereby minimizing strain along one of the in-plane directions. Under tensile strain, the lattice mismatch can be minimized by orienting the $c^+$ axis along the growth direction leading to an $a^-a^-c^+$ pattern. Indeed, this strain dependence of the in-phase axis has been predicted with density functional theory. For example, calculations of LaMnO$_3$ and CaTiO$_3$ reveal the $a^-a^+c^-$ pattern to be favorable under compressive strain and under tensile strain of less than 1~\% and 1.5~\%, respectively.\cite{Eklund09,LeePRB13} Similarly, the $a^-a^+c^-$ pattern was predicted to minimize energy in LaVO$_3$ in compressive strain.\cite{Sclauzero15} Under tensile strain above these values, the $a^-a^-c^+$ pattern becomes the lower energy structure. However, the energy differences between the two structural variants can be small; for example, first-principles calculations of strained LaVO$_3$ and many rare earth ferrites revealed minimal energetic preference between $a^-a^-c^+$ and $a^-a^+c^-$ structures under tensile strain.\cite{Sclauzero15,ZhaoJPCM14}

Our observation of an $a^-a^+c^-$ rotation pattern is consistent with previous experimental studies of epitaxial perovskites compressively strained to a non-$Pbnm$ substrate, including SrRuO$_3$/STO,\cite{ChangPRB11,LuPRB13,Vailionis08,ZiesePRB10} Pr$_{0.7}$Sr$_{0.3}$MnO$_3$/LAO,\cite{Mercey2000} LaVO$_3$/STO,\cite{Rotella2012} LaFeO$_3$/STO,\cite{Seo2008} GdTiO$_3$/STO/LSAT,\cite{Zhang2013} GdTiO$_3$/SrLaGaO$_4$,\cite{Grisolia14} Pr$_{0.5}$Ca$_{0.5}$MnO$_3$/LAO,\cite{Haghiri-Gosnet00} and La$_{0.9}$Sr$_{0.1}$MnO$_3$/STO.\cite{Vigliante01} The same structural variant has been reported in some films under small magnitudes of tensile strain, such as CaRuO$_3$/LSAT (0.55~\% tensile)\cite{Proffit2008} and PrVO$_3$/STO (0.5~\% tensile).\cite{Copie2013} In many of these studies, a mixture of $a^-a^+c^-$ and $a^+a^-c^-$ patterns was observed.\cite{Rotella2012,Seo2008,Mercey2000,Proffit2008,Copie2013}

There have also been reports of the $a^-a^-c^+$ pattern in films under tensile strain, especially in heterojunctions with larger than a 1~\% lattice mismatch. These studies include Pr$_{0.5}$Ca$_{0.5}$MnO$_3$/STO (2.3~\% tensile),\cite{Prellier2000} NdNiO$_3$/STO (2.6~\% tensile),\cite{Tung13} and La$_{0.7}$Ca$_{0.3}$MnO$_3$/STO (1~\% tensile).\cite{Andres03} A mixture of all three orientations was reported in CaMnO$_3$/LAO (2.3~\% tensile).\cite{Gunter12}

  \begin{figure}
\includegraphics[width=2.7 in]{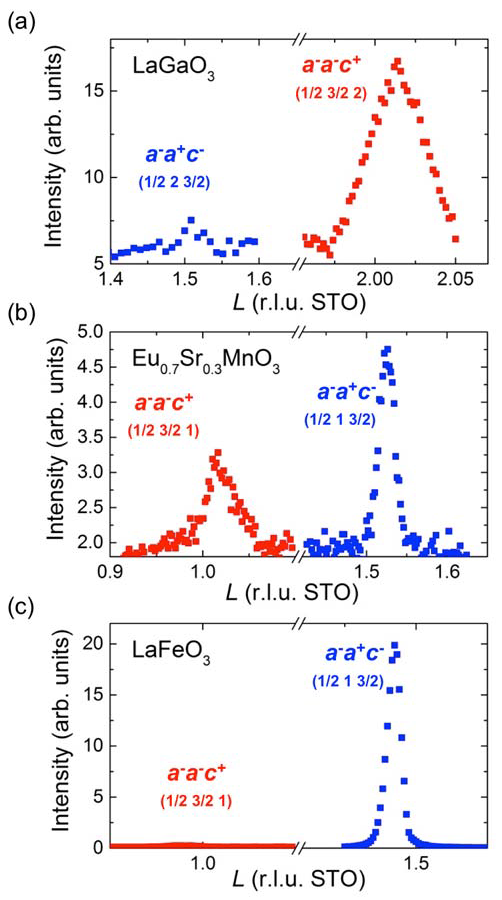}
\caption{(color online)  Measured ($\frac{1} {2}$ $\frac{3} {2}$ $n$) and ($\frac{1} {2}$ $n$ $\frac{3} {2}$) peaks for (a) LaGaO$_3$/STO, (b) Eu$_{0.7}$Sr$_{0.3}$MnO$_3$/STO, and (c) LaFeO$_3$/STO films, where $n = 2$ for (a) and $n = 1$ for (b) and (c). }
\label{fig:Fig7}
\end{figure}

To gain further insight into the in-phase axis orientation in films under moderate tensile strain (between 0 - 2~\%), we have measured the half-order peaks from LaGaO$_3$/STO (0.5~\% tensile) and Eu$_{0.7}$Sr$_{0.3}$MnO$_3$/STO (1.5~\% tensile). A survey of the half-order peaks indicate that both films retain the $Pbnm$-type rotation pattern that is found in bulk compounds. As shown in Fig. 7(a), the LaGaO$_3$ film is predominately $a^-a^-c^+$ oriented, which accounts for 94~\% of the sample volume compared to 6~\% for $a^+a^-c^-$ and $a^-a^+c^-$ domains as determined from intensity analysis of the half-order peaks. In contrast, the Eu$_{0.7}$Sr$_{0.3}$MnO$_3$ film is comprised of 65~\% $a^+a^-c^-$ and $a^-a^+c^-$ domains and 35~\% $a^-a^-c^+$ domains, based on the half-order peaks shown in Fig. 7(b). Finally, Fig. 7(c) shows half-order peaks measured from LaFeO$_3$/STO (-0.8~\% compressive strain) revealing over 99~\% of the film volume consists of $a^+a^-c^-$ and $a^-a^+c^-$ domains as expected for the film under compressive strain.

  \begin{figure}
\includegraphics[width=3.4 in]{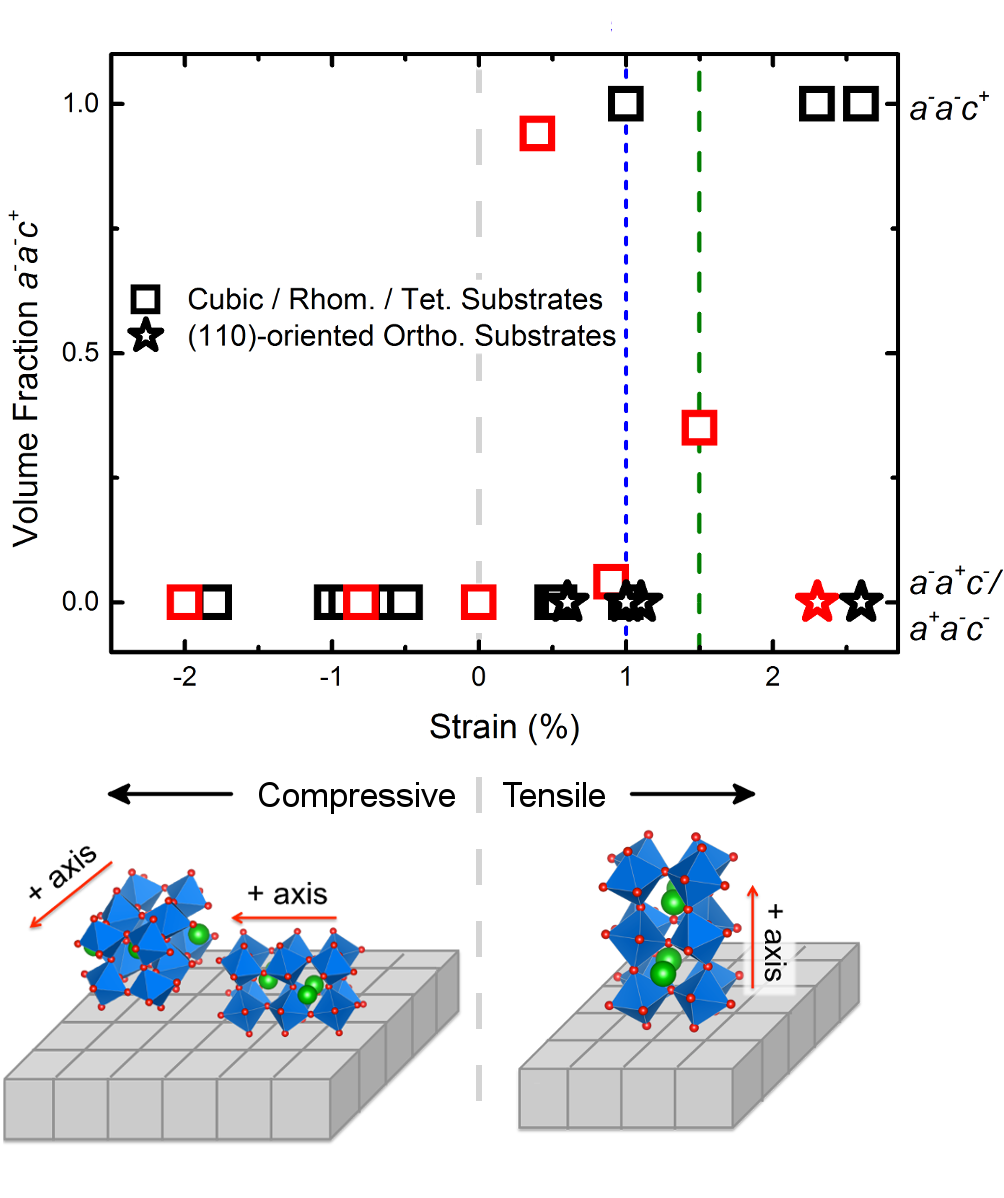}
\caption{(color online)  Compilation of experimental results reported in this study (red symbols) and previously published experimental work referenced in the main text (black symbols) displaying the orientation of the in-phase axis for $Pbnm$-type films. Films grown on non-$Pbnm$ substrates are indicated by squares, while films grown on (110)-oriented $Pbnm$ substrates are indicated by stars. The y-axis indicates the approximate volume fraction of the film that is $a^-a^-c^+$; a value of 0 indicates $a^+a^-c^-$ and/or $a^-a^+c^-$. The blue dotted and green dashed vertical lines signify the strain state at which a transition from $a^+a^-c^-$ to $a^-a^-c^+$ was predicted  in density functional calculations for LaMnO$_3$ and CaTiO$_3$.\cite{Eklund09,LeePRB13}}
\label{fig:Fig8}
\end{figure}

We compile our experimental results with those previously reported from both experiment and density functional theory in Fig. 8 to provide a comprehensive view of how the in-phase rotation axis responds to strain in films. From these results, three main conclusions can be drawn regarding the structural orientation of films on non-$Pbnm$ substrates. First, compressive strain leads to $a^-a^+c^-$ and/or $a^+a^-c^-$ structures. Second, large magnitudes of tensile strain ($> 2$~\%) promote $a^-a^-c^+$ structures. Finally, under moderate values of tensile strain (0 - 2~\%) films can exhibit any of the three in-phase orientations and in many cases considerable volume fractions of all three orientations. Within this strain range, it remains an open question regarding what factors, such as material chemistry or $B-B$ distances found in the bulk structure, determine the rotation pattern orientation of films. A table containing sample compositions and references for the data presented in Fig. 8 is given in the supplemental materials.

\subsection{Films on $\pmb{Pbnm}$ substrates} 
As discussed in the introduction, design strategies for realizing hybrid improper ferroelectrics and polar metals in short-period superlattices rely on the $A$-site ordering along the same direction as the in-phase rotation axis. This requires that the superlattices exhibit the $a^-a^-c^+$ structure. Based on Fig. 8, it is clear that such superlattices, when grown on cubic or rhombohedral substrates, must be under significant tensile strain to realize the correct orientation. However, the substrates most commonly used to induce large values of tensile strain are the rare earth scandates,\cite{SchlomMRS14} such as DyScO$_3$ and GSO, compounds that exhibit the $Pbnm$ structure.\cite{Liferovich04} In films grown on these substrates, the structural coupling between the film and substrate leads to an imprinting of the substrate in-phase axis orientation into the film. This imprinting effect is observed in the EuFeO$_3$/GSO films described here, and has also been reported in other papers detailing heteroepitaxial growth of $Pbnm$-type films under tensile strain on $Pbnm$-type substrates.\cite{Proffit2008,Kan2013,Aso14b,Biegalski2015} These results from films on (110)-oriented $Pbnm$ substrates, in which the in-phase axis within the substrate is perpendicular to the growth direction, are also plotted in Fig. 8 illustrating the substrate-induced structural coupling effect. The primacy of substrate imprinting over strain in determining the in-phase rotation axis points to growth on (001)-oriented $Pbnm$-type substrates as the most promising means to ensure $a^-a^-c^+$ behavior in perovskite films and superlattices.

Finally, it should be noted that we do not find an indirect imprinting effect on the in-phase axis from LAO, which exhibits an $a^-a^-a^-$ pattern, into $Pbnm$-type films. Here one may expect that the octahedral connectivity can be better maintained if the film takes on the $a^-a^-c^+$ pattern that would retain coherence of the out-of-phase axes within the epitaxial plane at the interface. However, in both our experimental results and those previously reported, such behavior is not found. This suggests that direct imprinting of the in-phase axis from a $Pbnm$ substrate provides deterministic control of the structural orientation while indirect imprinting from a rhombohedral substrate does not.

\section{Conclusions}

In summary, we report on the octahedral rotation patterns of strained EuFeO$_3$ and other $Pbnm$-type perovskite films.  We observe a mixed $a^-a^+c^-$ and $a^+a^-c^-$ rotation pattern when grown on cubic or rhombohedral substrates, in EFO films under strain states ranging from 2\% compressive to 0.9~\% tensile. In contrast, EuFeO$_3$ grown on orthorhombic GSO (110) exhibits a uniform $a^-a^+c^-$ orientation matching that of the substrate. To better understand the universality of this behavior, we have also measured LaGaO$_3$/STO, LaFeO$_3$/STO, and Eu$_{0.7}$Sr$_{0.3}$MnO$_3$/STO and compiled previously reported structural data from $Pbnm$-type films. The totality of the results indicates that compressive strain results in $a^-a^+c^-$ and $a^+a^-c^-$ patterns; moderate tensile strain can result in $a^-a^+c^-$, $a^+a^-c^-$, and/or $a^-a^-c^+$ structures; and large values of tensile strain ($>$ 2~\%) tends to favor $a^-a^-c^+$. However, films under large tensile strain on $Pbnm$-type substrates exhibit the same rotation pattern as that of the substrate, indicating that substrate imprinting of the in-phase axis offers a more robust means for deterministically controlling the rotation pattern compared to epitaxial strain. We anticipate that this work will enable more efficient experimental pursuits of recently predicted rotation-induced phenomena, such as hybrid improper ferroelectricity and non-centrosymmetric metals.

\section{Acknowledgements}
We thank Christian Schlep\"utz for assistance with the diffraction measurements. We are grateful to James Rondinelli and Craig Fennie for useful discussions. Work at Drexel was supported by the National Science Foundation (DMR-1151649).  Use of the Advanced Photon Source was supported by the U. S. Department of Energy, Office of Science, Office of Basic Energy Sciences, under Contract No. DE-AC02-06CH11357.

\bibliographystyle{apsrev-nourl}


\end{document}